\documentclass[fleqn,usenatbib,useAMS]{mnras}

\usepackage{graphicx}
\usepackage{lscape}   
\usepackage{color}
\usepackage{amsmath}
\usepackage{upgreek}

\title[The detection of strongly-lensed submillimetre galaxies]{The detection of strongly-lensed submillimetre galaxies}
\author[Chris Sedgwick et al.]
{Chris Sedgwick$^{1}$,            
Stephen Serjeant$^{1}$,        
Charles Weiner$^{1}$ \\\\                    
$^{1}$School of Physical Sciences, The Open University, Milton Keynes MK7 6AA, UK\\
}

\begin{document}

\date{Draft 6:  updated 26 March 2024}

\pagerange{\pageref{firstpage}--\pageref{lastpage}} \pubyear{2024}

\maketitle

\label{firstpage}

\begin{abstract}
We present predictions of
the number and properties of strongly-lensed submillimetre galaxies, based on an adaption of the physically-motivated {\sc LensPop} model  
covering galaxy-galaxy strong lensing by elliptical galaxies,  
which successfully predicted optical and near-infrared lenses. For submillimetre-luminous lensed galaxies, the most efficient observational selection identifies sources with high fluxes ($S_{500\upmu\textrm{m}}>80$\,mJy), where lensed sources outnumber bright unlensed sources; several hundred candidates from 
 {\it Herschel} surveys have been identified, and confirmed by follow-up observations. We have tested our model against these observations. The model predicts an all-sky number density of 0.09$\pm$0.05\,deg$^{-2}$ (in absolute numbers, 3\,600$\pm$1\,800) of bright lensed galaxies detectable by this method. Observations show considerable variation in sky density between fields, 0.08 - 0.31\,deg$^{-2}$. Predictions of redshift and magnification distributions are comparable to observations, although the model appears to under-predict lenses at the highest magnifications ($\mu>20$). We predict that the apparent AB magnitudes at visible wavelengths of the foreground lenses will be as faint as 28, whereas 
 observations typically reach $\sim23$, implying that some apparently unlensed bright submillimetre galaxies may have lensing galaxies below this detection limit. 
For fainter lensed galaxies, the model predicts over 130\,000 systems with flux $S_{500\upmu\textrm{m}}>10$\,mJy across the sky, of which $\sim3\,400$ remain be be discovered in the {\it Herschel} catalogues. We also predict that {\it Euclid} should be able to detect 
some 25\,000 lensed submillimetre galaxies that are VIS-band  
\textquoteleft dropouts\textquoteright ~- detectable in the near-infrared but not at optical wavelengths.

\end{abstract}


\section{Introduction}\label{sec:intro}

Dusty star-forming galaxies in the early universe are detectable in the submillimetre, since their optical/UV emission is reprocessed by dust into the far-infrared and then redshifted into the submillimetre region. These galaxies were first discovered with the Sub-millimetre Common User Bolometer Array 
(SCUBA) instrument at the James Clerk Maxwell Telescope (JCMT) at $850\,\upmu$m \citep{Smail1997,Hughes1998,Barger1998}. The number of submillimetre sources discovered in ground-based blank-field and cluster lens JCMT surveys has been increasing with advances in bolometer array instrumentation \citep[e.g. the SCUBA-2 Cosmology Legacy Survey][ covering $\sim5$\,deg$^2$ and  detecting $\sim$3\,000 sources]{Geach2017}. 

Strong gravitational lensing enables the detection of fainter and more distant submillimetre galaxies due to its angular and flux magnification of the background galaxy, enabling the study of star formation at high redshifts in more detail than would otherwise be possible. In addition, it enables the study of the substructure of dark matter halos in the foreground lens, helping to constrain the properties of dark matter, particularly in large samples. 
Although these lensed systems are rare objects, a technique for identifying them by their excess flux in wide-field catalogues was demonstrated by \cite{Negrello2010} and has led to the discovery of several hundred candidate lensed galaxies so far. 

The  {\it Herschel} Space Telescope \citep{Pilbratt2010} provided several  wide-field surveys at $500\,\upmu$m, e.g.: 
the {\it Herschel} Astrophysical Terahertz Large Area Survey \citep[H\nobreakdash-ATLAS;][]{Eales2010} covered 600\,deg$^2$,
the {\it Herschel} Multi-tiered Extragalactic Survey \citep[HerMES;][]{Oliver2012} covered $380$\,deg$^2$, and
the {\it Herschel} Stripe\;82 Survey \citep[HerS;][]{Viero2014} covered $79$\,deg$^2$.
 
The technique is a simple one: using early data from the H\nobreakdash-ATLAS survey, \cite{Negrello2010} showed that due to the steep drop in number counts of unlensed galaxies at $500\,\upmu$m, virtually all sources with flux $S_{500\upmu\textrm{m}}>100$\,mJy were strongly lensed, after other clearly-recognisable sources such as blazars and low-redshift spiral galaxies were removed. Later work identified further lensed submillimetre galaxies with {\it Herschel} data using this technique, such as  \cite{Wardlow2013}, \cite{Bussmann2013}, \cite{Nayyeri2016} and \cite{Negrello2017}. Recent work in the Bright Extragalactic ALMA Redshift Survey \citep[BEARS;][]{Urquhart2022} has used radio observations to confirm 71 submillimetre lenses in the H\nobreakdash-ATLAS SGP field which had been identified as lens candidates using the criterion $S_{500\upmu\textrm{m}}>80$\,mJy. The proportion of sources which are strongly lensed drops sharply below 100 mJy, but is still significant at 80 mJy particularly if an additional criterion such as a redshift cutoff is used \citep{Bakx2018}. This is the criterion we shall be using for most of our model predictions of lenses in this paper.

Ground-based wide-field surveys have also been made at millimetre wavelengths, such as a 2\,500 square degree survey \citep{Reuter2020} with the South Pole Telescope \citep[SPT;][]{Carlstrom2011}, and surveys of 455 and 840 square degrees \citep{Marsden2014,Gralla2020} with the Atacama Cosmology Telescope \citep[ACT;][]{Swetz2011}. The two ground-based surveys used a variation of the Negrello technique adapted for longer wavelengths. The SPT survey identified lens candidates based on flux measurements initially at 1.4~mm using the criterion of $S_{\rm 1.4\,\textrm{mm}}>20$\,mJy \citep{Viera2013}. This work was refined with observations by the Large Apex BOlometer CAmera (LABOCA; \citep{Siringo2009} using $S_{870\,\upmu\textrm{m}}>25$\,mJy \citep{Reuter2020}, who estimated that lensed sources accounted for $\sim$74\% of all sources identified with this criterion. The ACT survey used the criterion of $S_{\rm 1.4\textrm{mm}}>14$\,mJy \citep{Marsden2014}.

Data from the {\it Planck} cosmic microwave background experiment were also used to identify candidates \citep{PlanckXXXIX}, a subset of which were followed up with dedicated {\it Herschel} imaging \citep{PlanckXXVII, Canamares2015} and a machine-learning approach to identify lens candidates \citep{Lammers2022}. Although  submillimetre surveys by {\it Planck} are much shallower than those of {\it Herschel}, they are all-sky surveys and at least the very brightest lensed submillimetre galaxies should be detected, particularly those with the highest magnification \citep{Trombetti2021,Kamienski2024}.

Confirmation of the candidates found by the Negrello technique requires follow-up observations, such as redshift measurement of the source and where possible of the lens, and/or morphological confirmation of lensing from high resolution imaging. To examine the other properties of the source often requires sub-mm/mm-wave/radio observations since high-redshift submillimetre galaxies are usually difficult to detect at optical and near-infrared wavelengths. Where this has been done, the method of identifying candidate lenses with the Negrello criterion or the variants mentioned above has been shown to yield a very high success rate of confirmed lenses \citep{Negrello2010,Negrello2014,Negrello2017,Wardlow2013,Bussmann2013,Viera2013,Nayyeri2016,Reuter2020,Urquhart2022,Borsato2023}.

In this paper, we present predictions of the number and properties of strongly-lensed submillimetre galaxies using the {\sc LensPop} model. This model was originally developed to predict strong lensing of optical or near-infrared galaxies \citep{Collett2015}. We have adapted the model for use with submillimetre galaxies. 

The original {\sc LensPop} model is described in Section \ref{sec:model}, and its adaption to submillimetre galaxies is described in Section~\ref{sec:submm}. Observations made of submillimetre lenses are described briefly in Section~\ref{sec:observations} and these are compared to our model predictions in Section~\ref{sec:comparisons}. These first sections deal with lensed sources with fluxes $>80$\,mJy; sources with lower flux, down to $10$\,mJy, are considered in Section \ref{sec:beyond_negrello}. A discussion of the results and the potential for future observations is given in Section \ref{sec:discussion}. 

This paper assumes cosmological parameter values of 
$H_0=70$\,km\,s$^{-1}$\,Mpc$^{-1}$, $\Omega_\textrm{M}=0.3$ and $\Omega_\Lambda=0.7$.

\section{The {\sc LensPop} model}\label{sec:model}

The {\sc LensPop} model used as the basis for the work presented in this paper was developed by \cite{Collett2015} to predict strong gravitational lensing of optical galaxies, and is available as open source software\footnote{https://github.com/tcollett/LensPop}. As described in that paper, the code was verified against observations of  lenses by the galaxy-scale search of the Canada-France-Hawaii Telescope Legacy Survey \citep[CFHTLS;][]{More2012,More2016,Gavazzi2014}. The paper predicted lenses discoverable by the Wide Field Survey of the {\it Euclid} Space Telescope \citep{Laureijs2011}, the Dark Energy Survey \citep[DES;][]{DarkEnergy2016} and the Legacy Survey of Space and Time on the {Vera C. Rubin Observatory} \citep[LSST;][]{Ivezic2008,Abell2009}. 

The model was later used to predict results for the Cosmic Evolution Survey \citep[COSMOS;][]{Scoville2007} and the {\it Euclid} Deep Field in \cite{Weiner2019}. A further prediction of 17\,000 lensed near-infrared galaxies discoverable by the {\it Nancy Grace Roman Space Telescope} \citep[formerly known as WFIRST;][]{Green2012} was presented in \cite{Weiner2020}.

The {\sc LensPop} model assumes strong lensing by elliptical galaxies, which are modelled as singular isothermal ellipsoids (SIEs).  A population of  a billion foreground galaxies is generated using redshift, stellar velocity dispersion, flattening, effective radius and absolute magnitude in each of three bands (r, i, z) as parameters. A simulated population of 34\,331 sources prepared for the LSST collaboration \citep{Connolly2010} is used as the background (potentially lensed) source population. The lensing cross-section of the foreground lens population is then projected onto the background population to generate an idealized set of lens systems (lens and source pairs). The model then uses the observing parameters of the particular survey being considered to identify a set of detectable strongly-lensed sources. 

Four criteria are used in the original model \citep{Collett2015} for accepting the lens as detectable: that the source is within the Einstein radius; that the image is resolved; that the tangential shearing of arcs is detectable, and that the signal-to-noise ratio is over 20 in at least one  wavelength  band.

\section{Adaption of model to submillimetre galaxies}\label{sec:submm}

To adapt the model for use in predicting lensing of submillimetre galaxies, we have made two major changes: firstly, a background source catalogue of submillimetre galaxies was needed to replace the source catalogue of optical galaxies originally used; secondly, the criteria used to detect the lenses were altered.

A mock submillimetre catalogue of 75 million sources was created, based on the study by \cite{Cai2013}\footnote{based also on private correspondence between CW and Z-Y Cai.} which allows for an estimate of number counts as a function of the (unlensed) flux at $500\,\upmu$m and redshift. For simplicity, we have used only submillimetre galaxies at $z>1$. The unlensed number density based on the data from \cite{Cai2013} is consistent with that published for submillimetre galaxies \citep[e.g.][]{Negrello2010}. The source density parameter needed in the model was calculated from the data in \cite{Cai2013} to be 0.011\,arcsec$^{-2}$. The  model also requires the source galaxy angular size, which we have estimated using $z\sim 1-3$ and $z>3$ measurements from fig. 6 of \cite{Ikarashi2015}.

The second change to the model was to the acceptability criteria. Of the four criteria used in the original model, only the requirement for the source centre to be within the Einstein radius (Equation \ref{equation:equation1}) is retained:
\begin{equation}\label{equation:equation1}
\theta^{2}_{\rm E}>x^2_\textrm{S}+y^2_\textrm{S}
\end{equation}
where $\theta_{\rm E}$ is the Einstein radius and ($x_\textrm{S}$, $y_\textrm{S}$) are the coordinates of the centre of the source with respect to the lens; and a new criterion is added:
\begin{equation}\label{equation:equation5}
S_{\rm 500\upmu\rm m}>80{\rm\,mJy}
\end{equation}
where $S_{\rm 500\upmu\textrm{m}}$ is the observed flux at $500\,\upmu$m. Model predictions for lower-flux sources 
will be considered in Section~\ref{sec:beyond_negrello}.

The adaption of the {\sc LensPop} model to submillimetre galaxies is discussed in more detail in \cite{Weiner2019}, which also shows that the predictions of the adapted model are not sensitive to variations in cosmological parameters, for fixed observed source counts. The adapted {\sc LensPop} code is available as open source software\footnote{https://github.com/chrissedgwick/LensPop\_submm}.

\begin{figure}
 \begin{center} 
 \resizebox{3.2in}{!}{\includegraphics{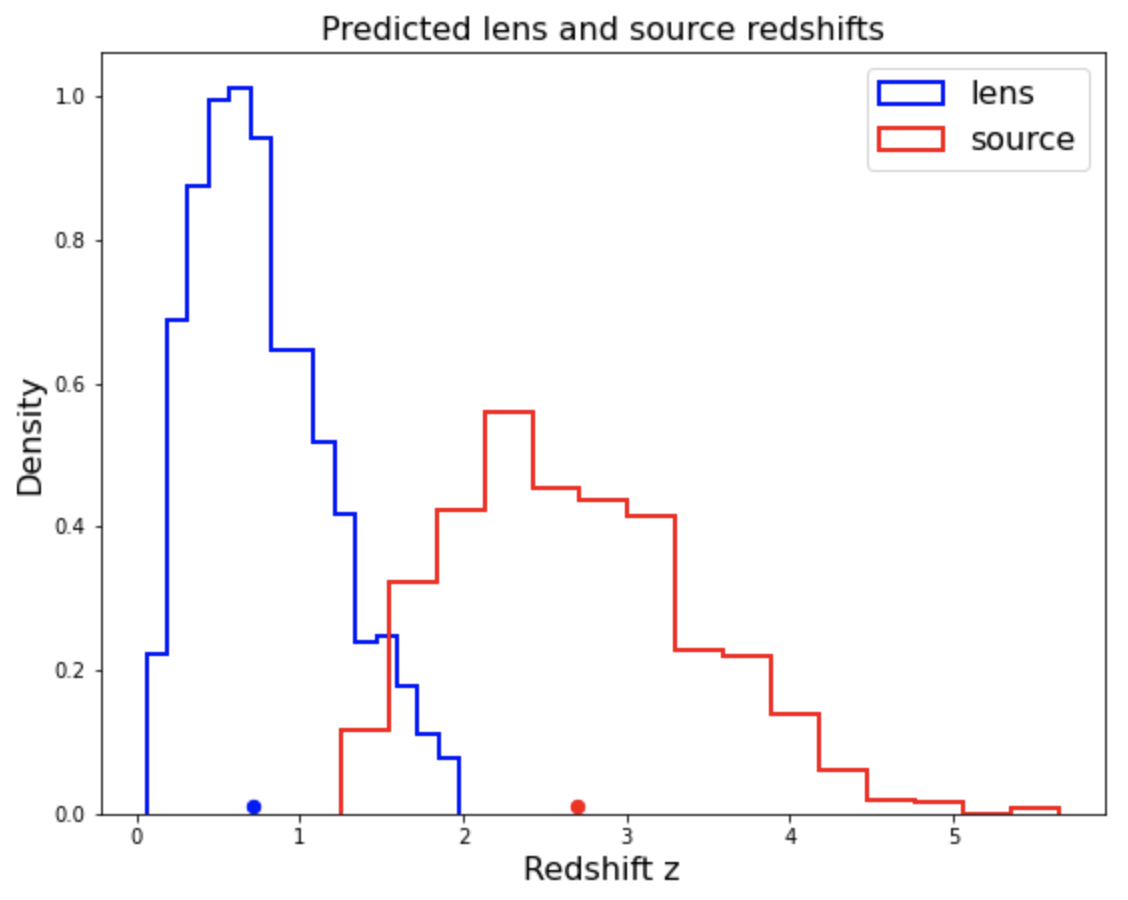}}     
\caption{{\sc LensPop} model predictions for lens and source redshifts of lensed submillimetre galaxies with S$_{500}>80$ mJy. Median redshifts are shown as points on the x-axis. Redshift predictions will be compared with observations in Section \ref{sec:comparisons}.}\label{fig:histograms}
\end{center}
\end{figure}

\begin{table*}

\caption{The surface density of confirmed strongly-lensed submillimetre galaxies from observations which use submillimetre flux criteria. Our model predicts a surface density of $0.09\pm0.05$\,deg$^{-2}$ (for galaxy-galaxy lenses only) using the $S_{500\upmu\rm m}>80$\,mJy criterion.} \label{table:comparisons}
\begin{center}
\begin{tabular}{|lllrrr|}
\hline

 Survey/field                               & Paper                  & Criterion                                     &   Area              &   Confirmed    & Surface density     \\
                                                   &                             &     (mJy)                                              &  (deg$^{2}$)    &   sources             & (deg$^{-2}$)  \\
\hline
\\

H\nobreakdash-ATLAS SGP field             &   Urquhart+ 2022        & $S_{500\upmu \rm{m}}\geq80$       & 262                  &  71                    &   $0.27\pm0.03$ \\
 
 H\nobreakdash-ATLAS NGP field            & Negrello+ 2017            & $S_{500\upmu \rm{m}}>100$    &  170                 &   14                   &   $0.08\pm0.02$     \\

H\nobreakdash-ATLAS equatorial fields   &  Negrello+ 2010, 2014  & $S_{500\upmu \rm{m}}>100$   &   16                  &   5                     & $0.31\pm0.14$  \\     
                                             &  Negrello+ 2017           & $S_{500\upmu \rm{m}}>100$      &  162                &   14                    & $0.09\pm0.02$     \\

HerMES  fields                   &  Wardlow+ 2013             & $S_{500\upmu \rm{m}}>100$     &   95                 &  9                       &  $0.09\pm0.03$     \\

H\nobreakdash-ATLAS+HerMES  fields    & Bussmann+ 2013         &$S_{500\upmu \rm{m}}>100$     &  395                &   25                     & $0.22\pm0.02$     \\

H\nobreakdash-ATLAS+HerMES  fields  & Calanog+ 2014              & $S_{500\upmu \rm{m}}>100$,$80$  &                  &     15                &                  \\

HerMES + HerS  fields         &  Nayyeri+ 2016            & $S_{500\upmu \rm{m}}>100$      &  372                &   13                     &           \\

All {\it Herschel} fields                & Borsato+ 2023             &   $S_{500\upmu \rm{m}}\geq80$   &   1\,000            &  65                      &   $>0.07$     \\\\

SPT                                     &  Vieira+ 2013                &   $S_{1.4 \rm{mm}}>20$         & 1\,300                &   38                    &  $>0.03$   \\
                                            &  Reuter+ 2020               & $S_{870\upmu \rm{m}}>25$       &  2\,500               &  81                     &  $>0.03$  \\\\
\hline

\end{tabular}
\end{center}
\end{table*}

\section{Observations of submillimetre lenses}\label{sec:observations}

The main studies of observations of submillimetre lenses using the Negrello criterion are listed in this section, and will be compared to our model predictions in the next section. These studies have used data from {\it Herschel} except for the South Pole Telescope observations which we have included for comparison.

 \cite{Bussmann2013} chose 30 candidates from the \mbox{H\nobreakdash-ATLAS} and HerMES catalogues for follow-up imaging of the sources with the Submillimeter Array (SMA) and optical and radio spectroscopy to identify the lens; redshifts of the sources were mostly already known. Confirmation was obtained that 25 of the sources were lensed.

 \cite{Wardlow2013} selected 13 candidates in the HerMES fields to follow up with various facilities, including submillimetre/radio interferometry with the SMA and the Jansky Very Large Array (JVLA), and near-infrared imaging with Keck-II and HST/WFC3. Confirmation of lensing was obtained for 9 sources.

\cite{Calanog2014} presented 87 candidates from H\nobreakdash-ATLAS and HerMES fields that were imaged in the near-infrared with Keck-Adaptive Optics and the {\it Hubble Space Telescope} (HST); clear lensing morphologies were observed for 15 candidates. Some of the other candidates were followed up in \cite{Borsato2023} (see below).

The study of \cite{Nayyeri2016} had 77 candidates that were chosen in the HerMES and HerS fields using the Negrello criterion, and presented follow-up observations which confirmed a lens for 13 of these candidates. The follow-ups were made with interferometric spectroscopy with the Green Bank Telescope (GBT), the Combined Array for Research in Millimeter-wave Astronomy (CARMA), and  the Plateau de Bure Interferometer (PdBI) and in some cases with near-infrared imaging with Keck and the William Herschel Telescope (WHT).

\cite{Negrello2017} listed 50 candidates in the H\nobreakdash-ATLAS NGP and equatorial fields that were cross-matched to available imaging and spectroscopic data  for confirmation, and 28 sources were confirmed, comprising 14 in each field. The earlier papers, \cite{Negrello2010} and \cite{Negrello2014} dealt with the original five submillimetre lenses discovered.  

\cite{Urquhart2022} continued this work in the southern hemisphere, presenting  85 candidates in the H\nobreakdash-ATLAS SGP field that were followed up with radio observations with the Atacama Large Millimeter Array (ALMA) 12-meter and ACA arrays. These  observations found that many of the candidates were multiple sources, giving 142 individual sources. Spectroscopic confirmation was obtained for 71 of these sources.

It is usually difficult to identify the foreground lensing galaxy in studies of submillimetre lenses; \cite{Borsato2023} used snapshot observations taken over a period of several years with the HST on lens candidates (with $S_{500\upmu\textrm{m}} \geq 80$\,mJy) from the {\it Herschel} surveys, finding candidates for the foreground lenses in 130 cases. From these,  65 systems were confirmed to be lenses, 30 being new discoveries.  
 
Finally, although selection criteria are at different wavelengths, we also include comparison with results from the ground-based SPT, which completed their ALMA follow-up campaign of lensed submillimetre galaxies with \cite{Reuter2020}. A total of 81 lensed sources were confirmed.

\begin{figure*}
 \begin{center} 
 \resizebox{6.6in}{!}{\includegraphics{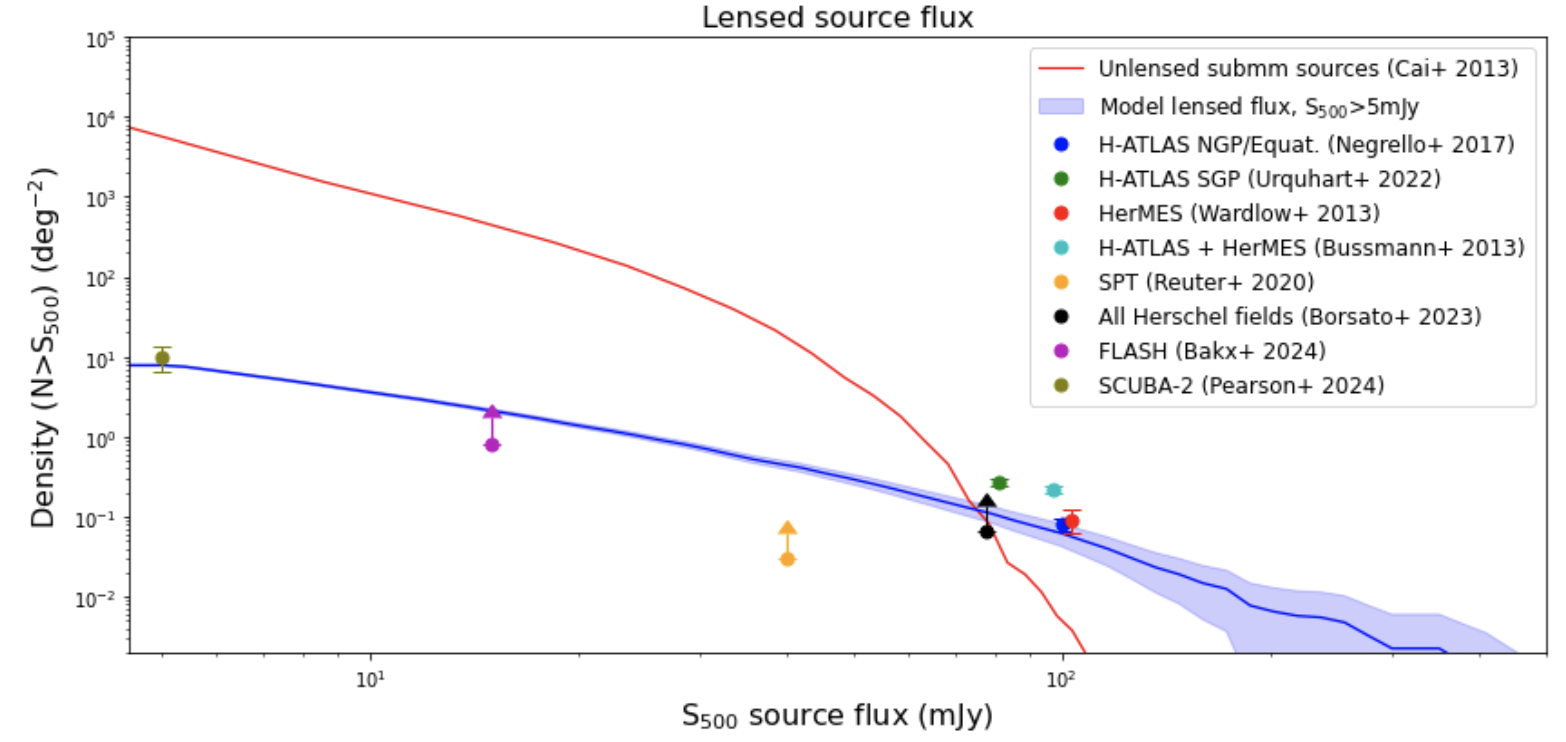}}     
\caption{The cumulative number density predicted by the model for the lensed source flux (in blue), with a $1\sigma$ error band, compared to observations of confirmed 
lenses to date with standard deviation bars assuming Poisson statistics. The vertical bars at $80$\,mJy and $100$\,mJy are displaced slightly for clarity. We omit compilations of unconfirmed lens candidates. 
Studies with additional selection criteria are shown as lower limits. Note that the model does not account for lensing by clusters or groups of galaxies, and systematic errors in the model assumptions may bias the lens number by tens of percent, as discussed in \protect\cite{Collett2015}. Also shown (in red) is the number density of the (unlensed) submillimetre sources in the model based on data in \protect\cite{Cai2013}. The figure suggests that unlensed sources may exceed lensed sources by an order of magnitude by $\sim$ 60 mJy.}\label{fig:number_counts2}
\end{center}
\end{figure*}

\section{Comparison with observations}\label{sec:comparisons}

\begin{table*}
\caption{{\sc LensPop} submillimetre galaxy predictions of redshifts, magnification, and source flux compared to observations from recent surveys.}\label{table:predictions}
\begin{center}
\begin{tabular}{|lrrrrrrrr|}
\hline
                                         &\multicolumn{2}{c|}{---Lens redshift ($z_\textrm{L}$)---}     & \multicolumn{2}{c|}{--Source redshift ($z_\textrm{S}$)--}  & \multicolumn{2}{c|}{---Magnification ($\mu$)---}  & \multicolumn{2}{c|}{--Lensed source flux S$_{500}/$mJy--}\\   
                                        &   median           &range~~                       &    ~~~~median    &   range~~                   &  ~~~~~median                &      range~~    &  ~~~~~median                &      range~~   \\                                         
 
\hline
\\
{\bf {\sc LensPop} model: } \\
$S_{500\upmu\textrm{m}}>100$\,mJy         &     0.69                  &  0.1 - 2.0                &        2.6~                   &  1.3 - 4.8               &   6.24~          &  2.2 - 32       &     125            & 100 - 492   \\
$S_{500\upmu\textrm{m}}>80$\,mJy         &     0.72                  &    0.1 - 2.0                &         2.7~                  &   1.3 - 5.0              &   5.65~          &  1.9 - 33      &     104              &  80 - 492   \\\\
\hline

{\bf Observed data}            \\\\
{\bf Herschel}\\

Bussmann+ 2013  (25)          &    0.6~                   &  0.2 - 1.4                &     2.5~             &   1.3 - 4.2                 &     6.9~                    & 1.2 - 15           &   203          &  101 - 344      \\
Wardlow+ 2013       (9)           &   0.6~                   &  0.4 - 1.4                &    2.8~               &   1.3 - 3.4                 &   $\sim9$~             & 1.5 - 23           &     140        &  110 - 249         \\
Nayyeri+ 2016        (13)          &   0.4~                   &  0.1 - 1.0                &   2.5~               & 1.2 - 5.2                    &           -                   &           -             &  164          &   121 - 718     \\
Negrello+ 2017       (28)          &   0.6~                    &  0.2 - 1.2               &   2.5~               &  1.0 - 4.2                  &   8.2~                      & 1.8 - 16           &    176         &    104 - 343     \\ 
Urquhart+ 2022       (71)         &  0.7~                     &  0.5 - 1.3               &   2.7~             & 1.4 - 4.5                     &   4.8~                      &  1 - 50             &     88         &  80 - 204           \\
Borsato+ 2023        (65)         &   0.6~                      & 0.1 - 1.4                &  2.6~                  &  1.0 - 5.2                    &  6.1~                              &  1.9 - 28           &   130              &  55 - 718        \\\\

{\bf Ground-based} \\
Reuter+ 2020            (81)        &       -                    &  -                             &  3.9~                  &  1.9 - 6.9                &    5.5~                 &   1.18 - 33           &      128           &   41 - 555   \\\\
\hline      
\end{tabular}
\end{center}
\end{table*}

The predictions of the model for various properties of the submillimetre lensing systems observable with {\it Herschel} are discussed in this section, and compared with the results of the observations with $S_{\rm 500\upmu\rm m}>80$ or $>100$\,mJy. 

\subsection{Galaxy group and cluster lenses}

In comparing the predictions to observations, it should be borne in mind that the model predicts only galaxy-galaxy lenses, whereas observations will  include galaxy group lenses and cluster lenses. Papers of observations have identified a number of these, but are not always able to distinguish them, so their number is unclear, but may be of the order of 10\% - 20\%. For example, \cite{Negrello2017} identified two group lenses and two cluster lenses out of 28 confirmed lenses; \cite{Bussmann2013} identified two clusters lenses out of 25 identified lenses; in the SPT observations, \cite{Spilker2016} found four lenses out of 47 to be galaxy group or cluster lenses.

\subsection{Surface number density} 

The model predicts that 3\,640 galaxy-galaxy lenses (all-sky) with flux $S_{\rm 500\upmu\rm m}>80$\,mJy will be detectable, giving a sky density of $0.09\pm0.05$\,deg$^{-2}$. 

The observed surface number density of submillimetre lensed systems for each field covered by studies reported to date is shown in Table~\ref{table:comparisons}. The surface density in the different fields ranges from $0.08\pm0.02$\,deg$^{-2}$ for the H\nobreakdash-ATLAS Northern Galactic Pole (NGP) field, to $0.27\pm0.03$\,deg$^{-2}$ for the Southern Galactic Pole (SGP) field and $0.31\pm0.14$\,deg$^{-2}$ for the earlier studies of the H\nobreakdash-ATLAS equatorial fields. The recent, larger study across all {\it Herschel} fields by \cite{Borsato2023} found a surface density of $0.07\pm0.01$\,deg$^{-2}$, but this only included sources whose foreground lenses were detected with HST. The observations suggest variability across the sky, although some of the fields cover a fairly small fraction of the whole sky. The observed variability is perhaps caused by cosmic variance, driven partly by lensing galaxies being preferentially biased towards richer environments. 

The number density predicted by the model of $0.09\pm0.05$\,deg$^{-2}$ is within the observed range, although nearer the lower point. This may be because, in part, group lenses and cluster lenses are not predicted by the model; also, as discussed below, it seems the model may under-predict the number of the highest-magnification sources.

The ground-based surveys, which used $870\,\upmu$m and 1.4\,mm criteria, are less sensitive than the {\it Herschel} surveys, and show a lower limit surface density of $0.03$\,deg$^{-2}$. All-sky ground-based surveys with equipment similar to the SPT should yield perhaps 1\,300 submillimetre lenses.

The  number count density of lenses by source flux is shown in Fig.\,\ref{fig:number_counts2}, which covers a range from $10$\,mJy upwards, and shows the density counts observed in the studies to date at fluxes of $100$\,mJy and $80$\,mJy. It also shows the count density for observations with the SPT \citep{Reuter2020} above $40$\,mJy, and with the FLASH project \citep[][see Section \ref{sec:beyond_negrello} below]{Bakx2024} above $15$\,mJy.

\begin{figure}
 \begin{center} 
   
  \resizebox{3.3in}{2.55in}{\includegraphics{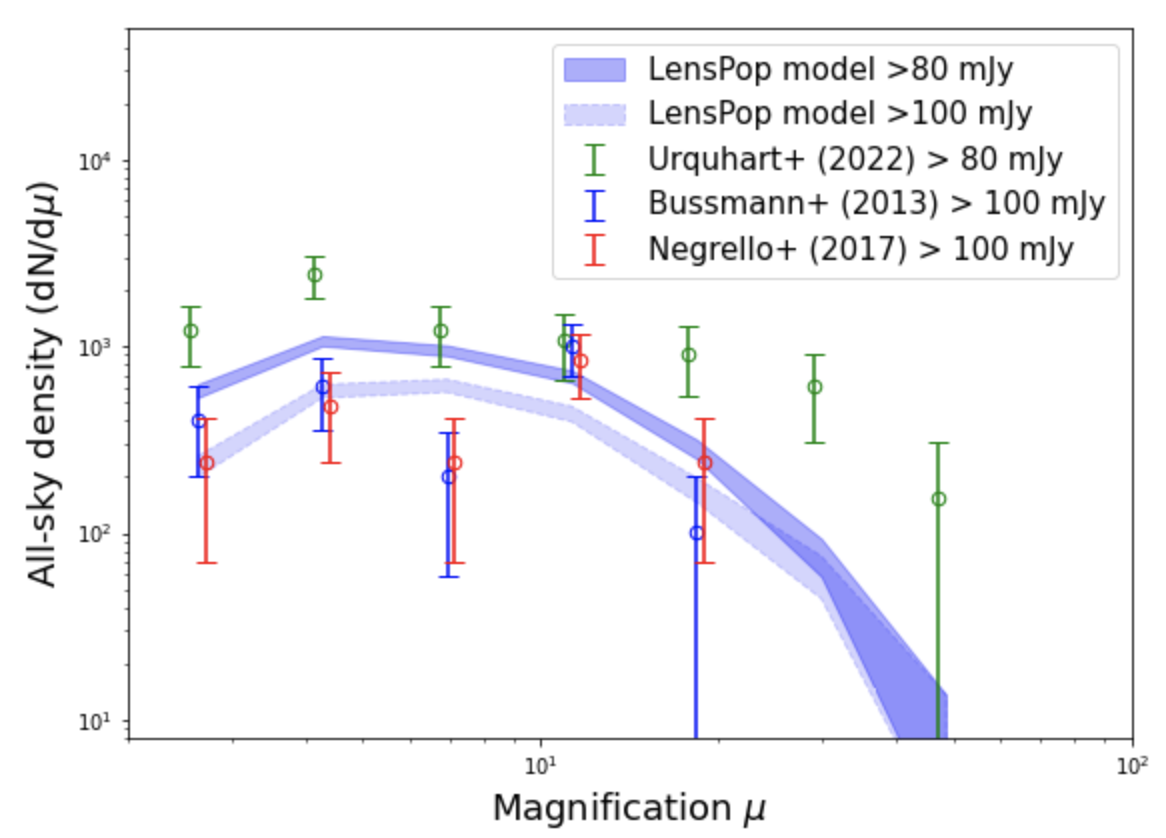}}  
   \resizebox{6.0in}{0.2in}{\includegraphics{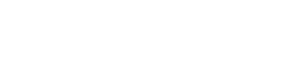}}   
   \resizebox{3.3in}{2.55in}{\includegraphics{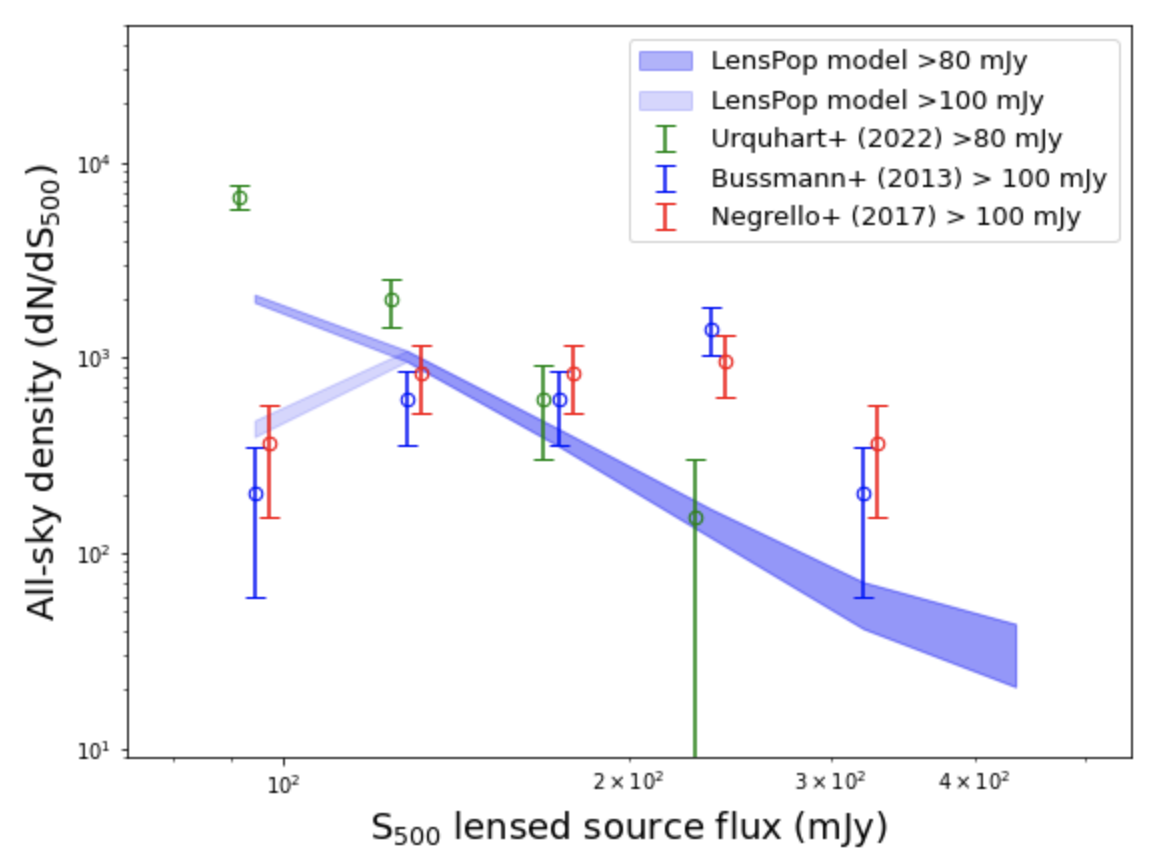}} 
    \resizebox{6.0in}{0.2in}{\includegraphics{white_space.png}}  
    \resizebox{3.3in}{2.55in}{\includegraphics{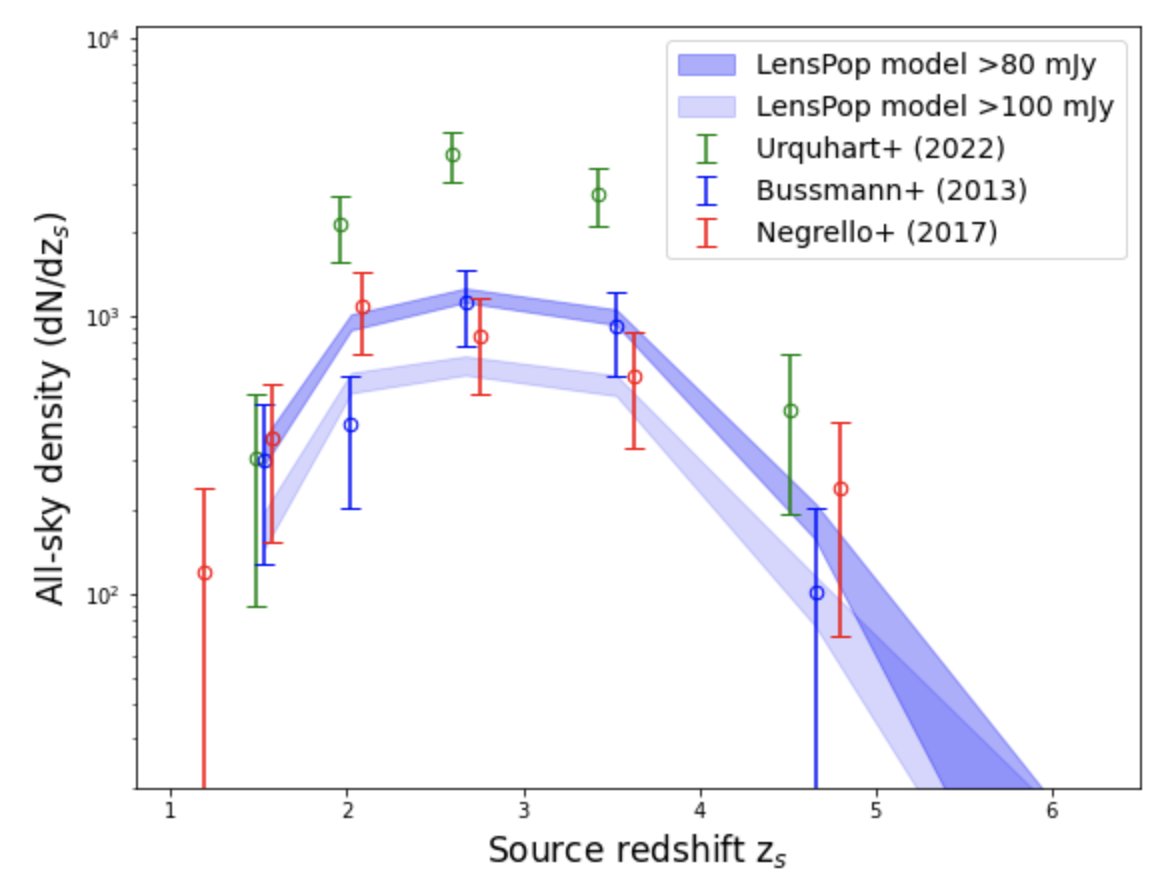}}       
\caption{Comparison of model predictions with observations. From top: (a) the magnification distribution for $\mu\ge2$; (b) lensed source flux; and (c) source redshifts. The vertical bars are displaced slightly for clarity. The model predictions are shown as a blue $\pm1\sigma$ band. The flux data for Urquhart observations is from \protect\cite{Bakx2018}. Poisson errors are assumed for the bin counts of the observations. }\label{fig:urquhart}
\end{center}
\end{figure}

\begin{figure}
 \begin{center} 
 \resizebox{3.2in}{!}{\includegraphics{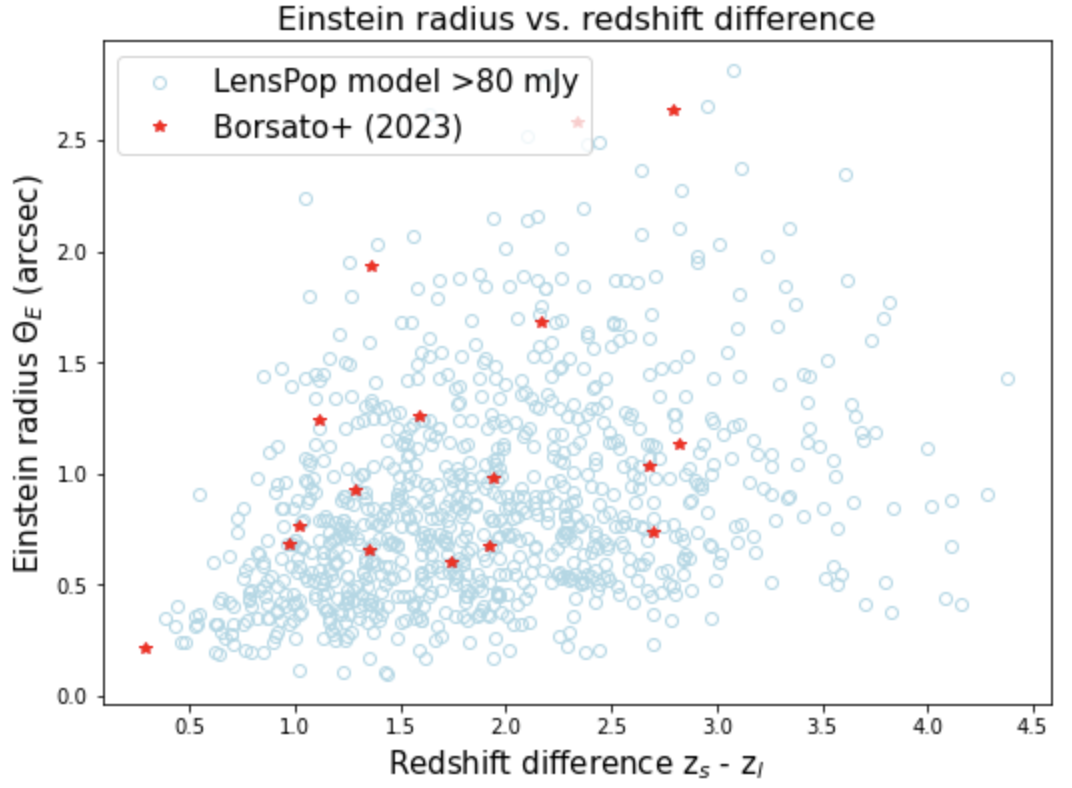}}     
\caption{Comparison between the 17 lenses in \protect\cite{Borsato2023} for which the Einstein radius and both redshifts are available and model predictions for $S_{500}>80$\,mJy.}\label{fig:einstein2a}
\end{center}
\end{figure}

\subsection{Redshifts}

For source redshifts, the model predicts a median of 2.65 and a mean of $z_\textrm{S}=2.72\pm0.01$ (standard deviation 0.74); for lens redshifts, the model predicts a median of 0.72 and a mean of $z_{\rm L}=0.77\pm0.01$ (standard deviation 0.40), using the $S_{500\upmu\textrm{m}}>80$\,mJy criterion. A histogram of the source and lens redshift distributions predicted by the model is shown in Fig.\,\ref{fig:histograms}. Redshifts from observations are compared to model predictions in Table~\ref{table:predictions}. 

For background source redshifts $z_\textrm{S}$, both the median and the range from the model match observations closely. Fig.\,\ref{fig:urquhart}(c) shows the number density by redshift: the results from \cite{Urquhart2022} are higher than the model predictions, but the other {\it Herschel} studies are reasonably close to the model.

The situation is different for the foreground lens redshifts: the model shows a range of $z_\textrm{L}$ from 0.06 to 2.0, whereas none of the observations reaches beyond $z_\textrm{L}$=1.4 (see Fig.\,\ref{fig:lens_predictions}). A similar result also shows up clearly in the range of lens apparent magnitudes, discussed below. 


\begin{figure*}
 \begin{center} 
  \resizebox{5.0in}{!}{\includegraphics{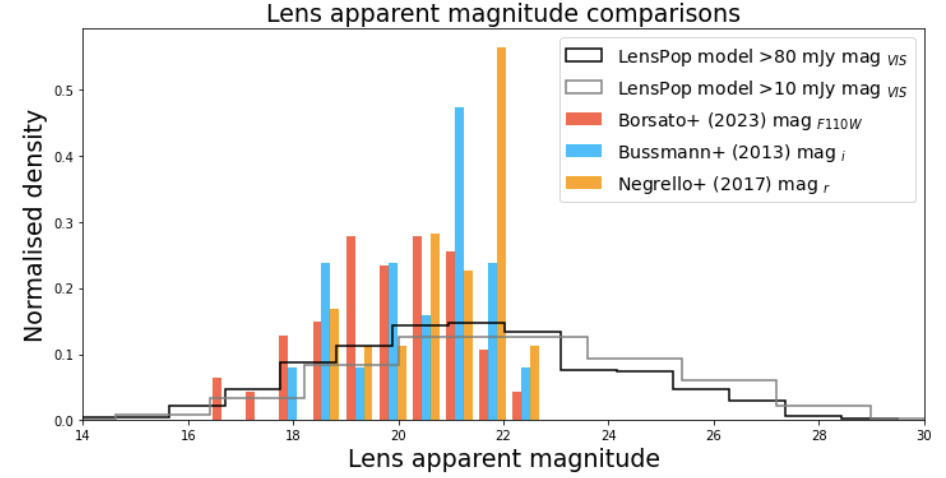}}  
\caption{ The model predicts that the apparent magnitude of many foreground lenses in submillimetre lens systems is well below that of observations to date. Note that the wavelength bands are different in each case: an average 'visible' bands for the model, SDSS r-band for \protect\cite{Negrello2017}, SDSS i-band for \protect\cite{Bussmann2013}, the HST F110W band for \protect\cite{Borsato2023} (all AB magnitudes).} \label{fig:magnitude2}
\end{center}
\end{figure*}

\subsection{Magnification}

For magnification, the model predicts a median of 5.65 and a mean of $\mu=7.10 \pm 0.08$ (standard deviation 4.96), using the $S_{500\upmu\textrm{m}}>80$\,mJy criterion.

The observed magnifications are shown in Table~\ref{table:predictions}; medians vary from $\mu=4.8$ in \cite{Urquhart2022} up to $\mu>8$ in \cite{Bussmann2013} and \cite{Wardlow2013}. The range of magnification also varies between observations, particularly at the high end, with \cite{Urquhart2022} reporting a maximum of $\mu=50$ (though with a large error bar). The {\sc LensPop} model predicts a maximum of $\mu=33$. The plots of magnification in Fig.\,\ref{fig:urquhart}(a) suggest that the model under-predicts magnification from around $\mu\sim20$ upwards. This may be due to increased optical depth obtained by galaxy cluster or group lenses, which are not covered in the model.

\subsection{Lensed source flux}

The (magnified) source flux comparisons are shown in Fig.\,\ref{fig:urquhart}(b). \cite{Urquhart2022} shows a lensed source flux distribution similar that predicted by the model, whereas the observations by \cite{Negrello2017} and \cite{Bussmann2013} show a density of high-flux sources a bit higher than the model, probably for the same reasons discussed above in relation to the results for number density and magnification.

\subsection{Einstein radius}

For the Einstein radius of the lens systems, the model predicts a mean of ${\rm \theta_{\rm E}}=0.86\pm0.49$\,arcsec (median 0.75\,arcsec), using the $S_{500\upmu\rm m}>80$\,mJy criterion.

\cite{Borsato2023} gives the Einstein radius for 23 of its sources (including data from previous studies of its sources), with values ranging from $0.215^{+0.006}_{-0.003}$\,arcsec to $2.631^{+0.007}_{-0.012}$\,arcsec. The observed values are compared to the model in relation to the difference $z_\textrm{S}-z_\textrm{L}$ in Fig.\,\ref{fig:einstein2a}, and are broadly consistent.

\subsection{Lens apparent magnitude}

Three papers have reported AB apparent magnitudes of the foreground lenses at optical or near-infrared wavelengths: \cite{Bussmann2013} at the SDSS  r-band, \cite{Negrello2017} at the SDSS i-band, and \cite{Borsato2023} at the HST F110W band.
Fig.\,\ref{fig:magnitude2} compares these observations with the model prediction for an average of SDSS r, i and z bands, referred to as VIS.\footnote{This follows the original {\sc LensPop} model, and roughly corresponds to the centre of the waveband observed by the {\it Euclid} VIS instrument.} 

Whereas none of the observations show lenses at magnitudes below $23$, the model predicts magnitudes down to $\sim28$, several orders of magnitude lower, which is perhaps unexpected.
These low-magnitude lenses often relate to sources at higher redshift and systems with lower Einstein radii  (see Fig.\,\ref{fig:area_zs}), making their detection difficult. 
 
\begin{figure*}
 \begin{center} 
 \resizebox{2.95in}{2.2in}{\includegraphics{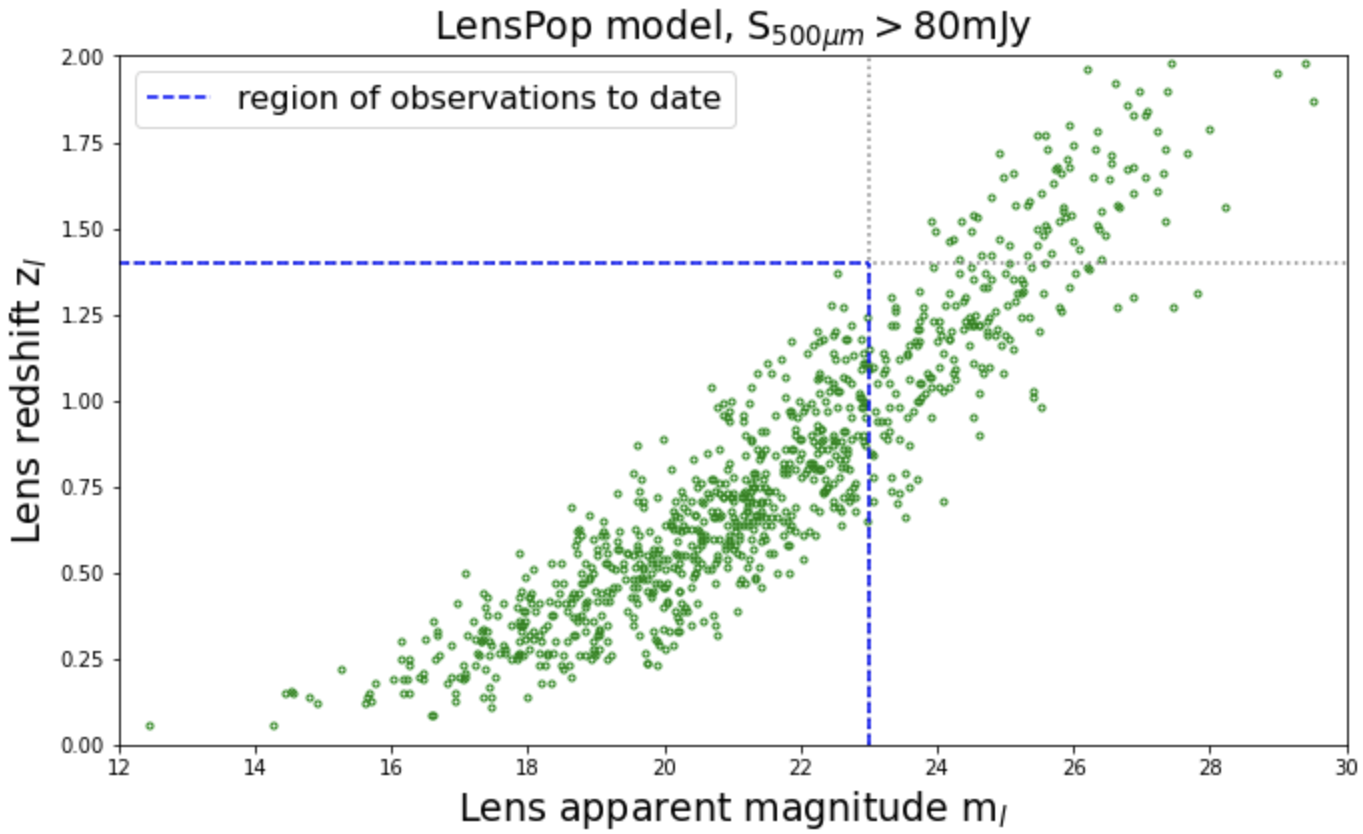}}     
     \resizebox{0.1in}{2.0in}{\includegraphics{white_space.png}}     
  \resizebox{3.45in}{!}{\includegraphics{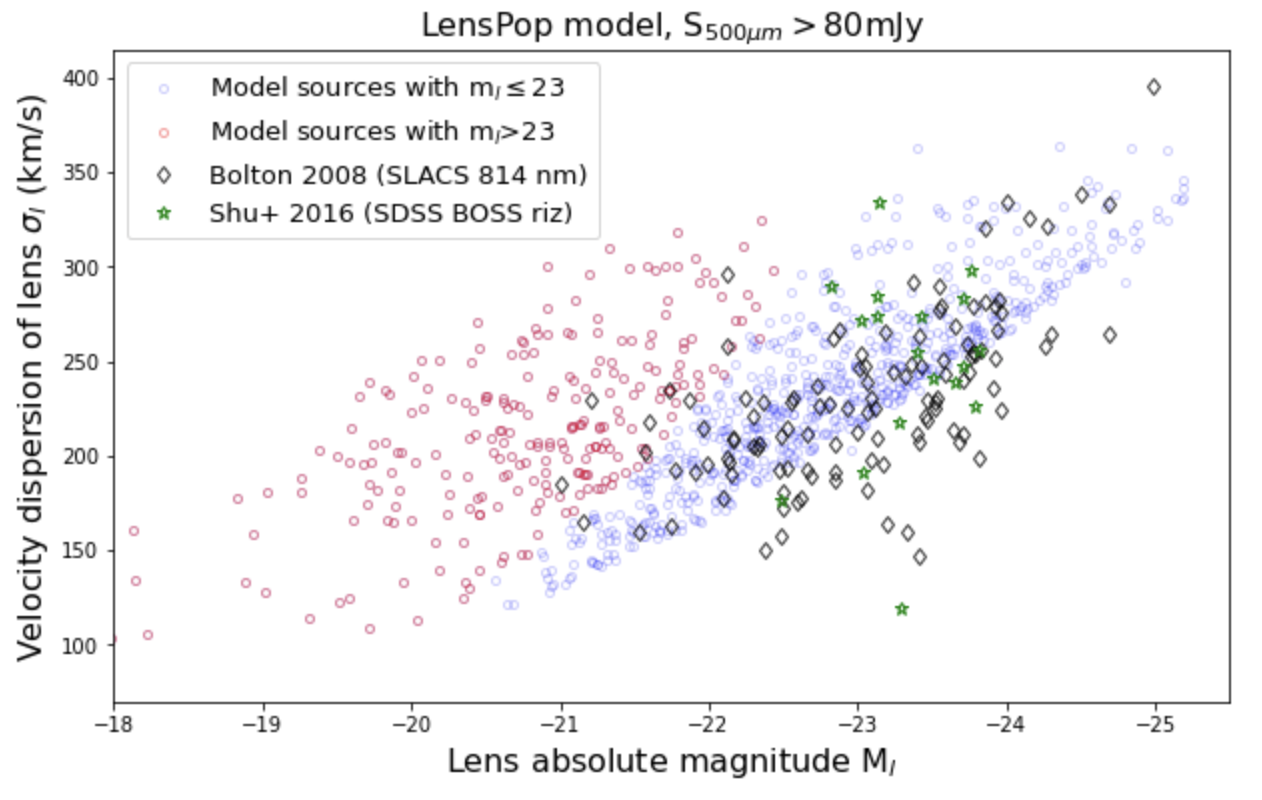}}  
\caption{The left-hand panel demonstrates observations to date under-estimate high-$z$, low-magnitude lenses. The simulted lens apparent magnitude in the VIS filter is plotted against lens redshift, and an approximate apparent magnitude limit corresponding to the observations in Fig.\,\ref{fig:magnitude2} is plotted as a vertical line. This corresponds to an effective lens redshift limit below 1.5, shown by the horizontal line. The right-hand panel shows the predicted Faber-Jackson relation, i.e. the relation between velocity dispersion (as a proxy for galaxy mass) and absolute magnitude.  The predicted low-apparent-magnitude 
lenses 
show a similar mass:luminosity relation to that observed for faint elliptical galaxies \citep{Bolton2008, Shu2016}, although the simulation does not reproduce the small number of low-mass lenses.
The luminosity for the BOSS sources is taken as an average of SDSS r,i,z wavelengths, like the model, in order to approximate the VIS filter; the wavelength for the luminosity for the SLACS data is 814\,nm. }\label{fig:lens_predictions}
\end{center}
\end{figure*}

 The Faber-Jackson mass-to-light ratios \citep{Faber1976} of the fainter lenses predicted by the model are similar to those observed for faint elliptical galaxies \citep{Bolton2008, Shu2016} (see the right-hand panel of Fig.\,\ref{fig:lens_predictions}), although some low-mass lenses which have been detected in these studies are not predicted by the model.


\subsection{Other properties}

\begin{figure*}
 \begin{center} 
 \resizebox{2.17in}{!}{\includegraphics{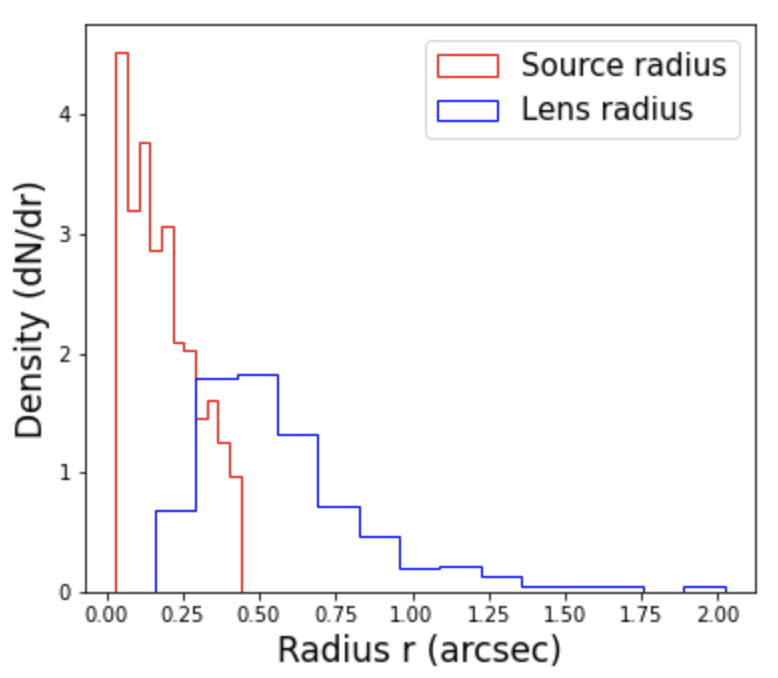}}     
  \resizebox{2.2in}{!}{\includegraphics{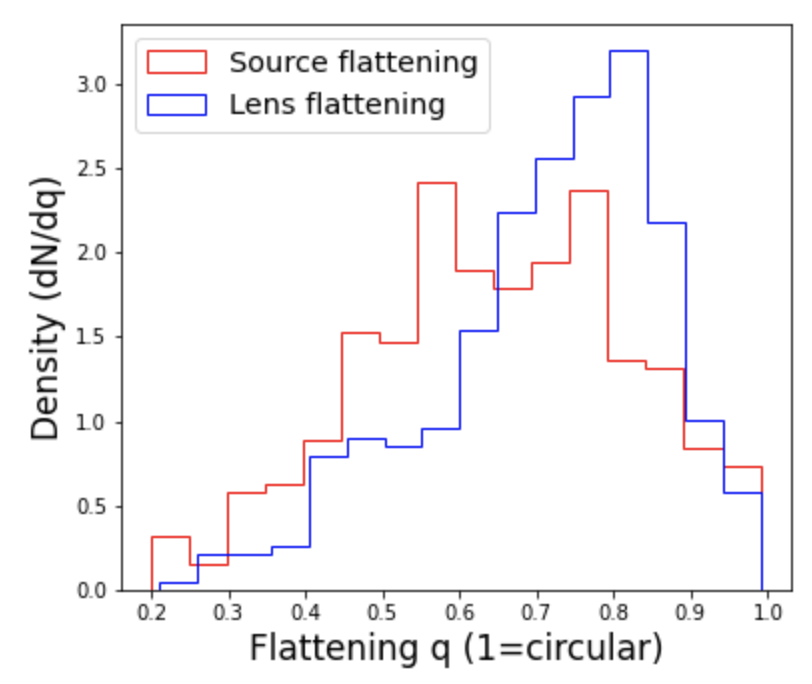}}    
    \resizebox{2.22in}{!}{\includegraphics{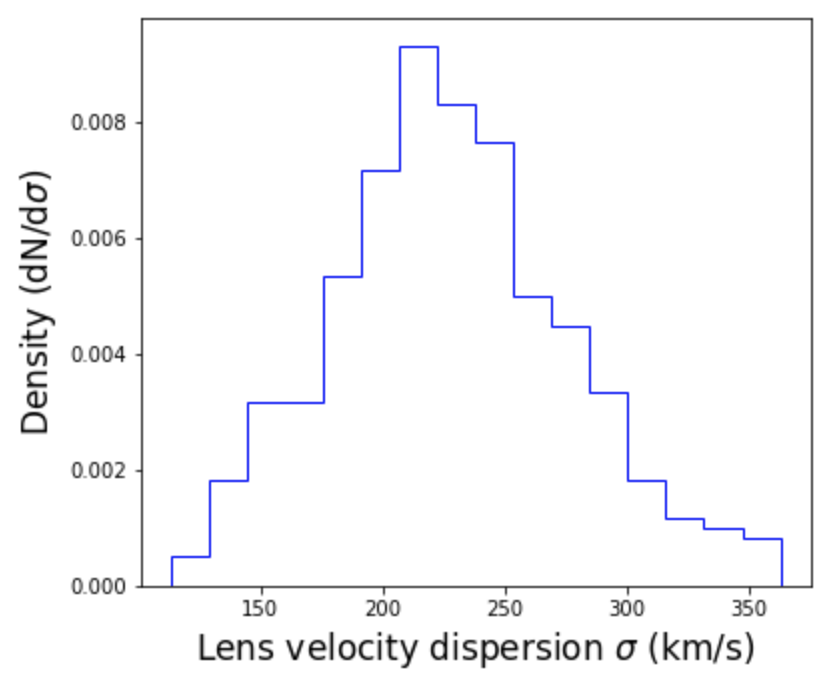}}  
\caption{A selection of other observable properties predicted by the {\sc LensPop} model, for $S_{500}>$80\,mJy. The flattening parameter q is the ratio of semi-major to semi-minor axes; the population has been truncated at q=0.2. (Note that more massive galaxies are more likely to be closer to spherical). The lens velocity dispersion function used in the model was derived from SDSS elliptical galaxies \citep{Choi2007} and it is assumed that it does not evolve with redshift.} \label{fig:other_props}
\end{center}
\end{figure*}

The model also predicts other properties not yet reported in observations, which are shown in Fig.\,\ref{fig:other_props}. 

Velocity dispersion (for the lenses) ranges from about $100 - 350$\,km\,s$^{-1}$ with a mean of $\sigma=225\pm50$\,km\,s$^{-1}$ (median $220$\,km\,s$^{-1}$). 

The radius and shape of both lenses and sources are also predicted: the source half-light radius has mean $0.19\pm0.11$\,arcsec (median $0.17$\,arcsec); the lens radius is typically under 1\,arcsec, only reaching out to just under 2\,arcsec. Most lenses are predicted to be close to circular, but sources are predicted to have a wider range of shapes, with source flattening (circular=1) of mean $0.65\pm0.18$.

\begin{figure*}
 \begin{center}     
   \resizebox{3.3in}{!}{\includegraphics{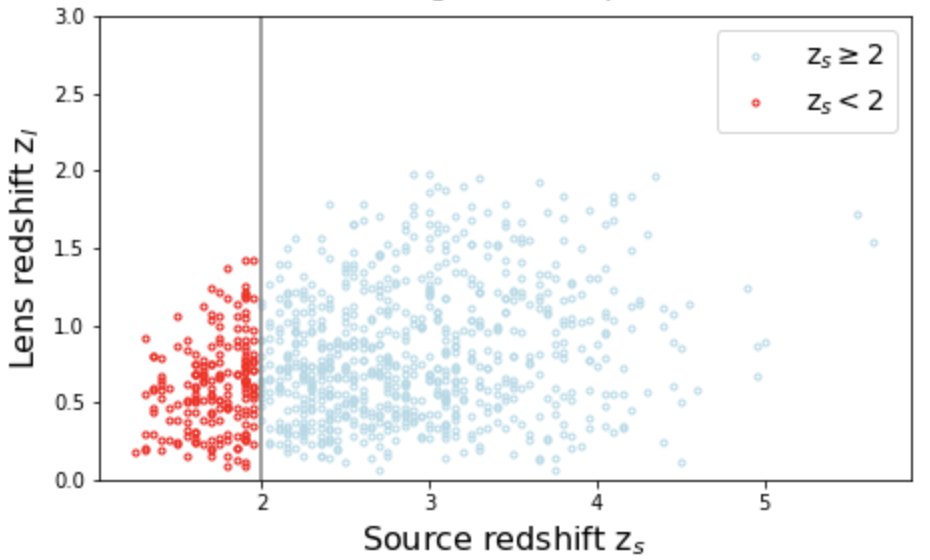}}     
   \resizebox{3.2in}{!}{\includegraphics{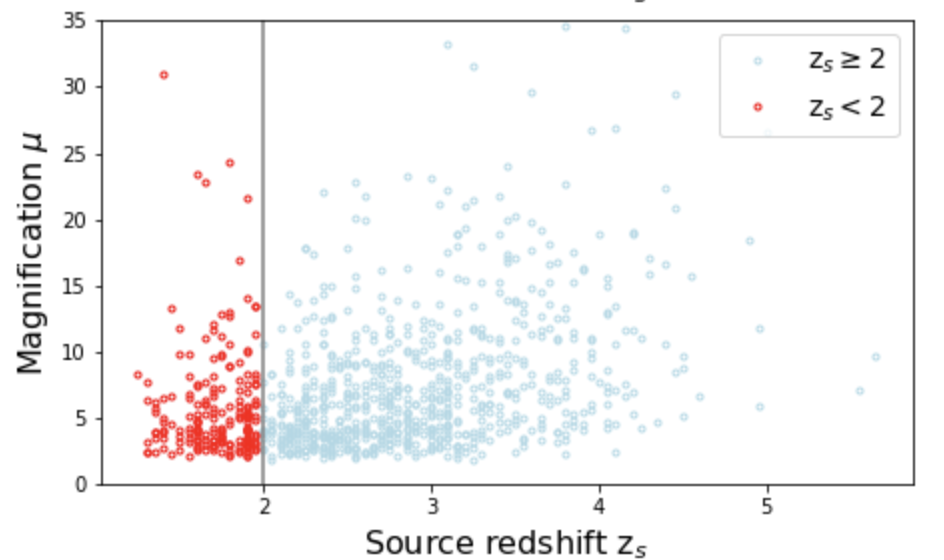}}    
\caption{The model (with $S_{500\upmu\rm m}>80$\,mJy) suggests that applying a cutoff at source redshift $z_\textrm{S}$=2 (as in \protect\cite{Bakx2018}) does not significantly affect the properties of the selection: examples showing effect for lens redshift and for magnification shown in this figure. }\label{fig:herbs}
\end{center}
\end{figure*}

\section{Lenses at lower source flux density}\label{sec:beyond_negrello}

\begin{table}
\caption{Predicted counts of strongly-lensed submillimetre galaxies by observed flux band.} \label{table:counts_by_flux}
\begin{center}
\begin{tabular}{|lrrr|}
\hline

Lensed flux    & \multicolumn{3}{c|}{-------- Predicted Count --------}      \\
   $S_{500\upmu\textrm{m}}$   &  Whole sky  & {\it Herschel} data         & Percent       \\
      (mJy)                   &                   &  1000\,deg$^2$   &      (\%)         \\
\hline
\\
Over 100        &  2\,096           & 52           &   1.5    \\
90-100            &    696            & 17           &  0.5      \\
80-90              &  884              & 21           &  0.6    \\
70-80              &  1\,420          & 35            &   1.0     \\
67-70              &  2\,220          &  56           &  1.6      \\
50-60               & 3\,468         &   87           &   2.5     \\
40-50               & 5\,768         &  144          &   4.2     \\
30-40               &  10\,764      &   269         &  7.9      \\
20-30               & 24\,952       &    624        &  18.2     \\
10-20               & 84\,844       &   2\,121      &  61.9     \\\\
Total                & 137\,112      &  3\,428       & 100      \\\\
\hline
\end{tabular}
\end{center}
\end{table}

\begin{figure*}
 \begin{center} 
  \resizebox{6.6in}{!}
  {\includegraphics{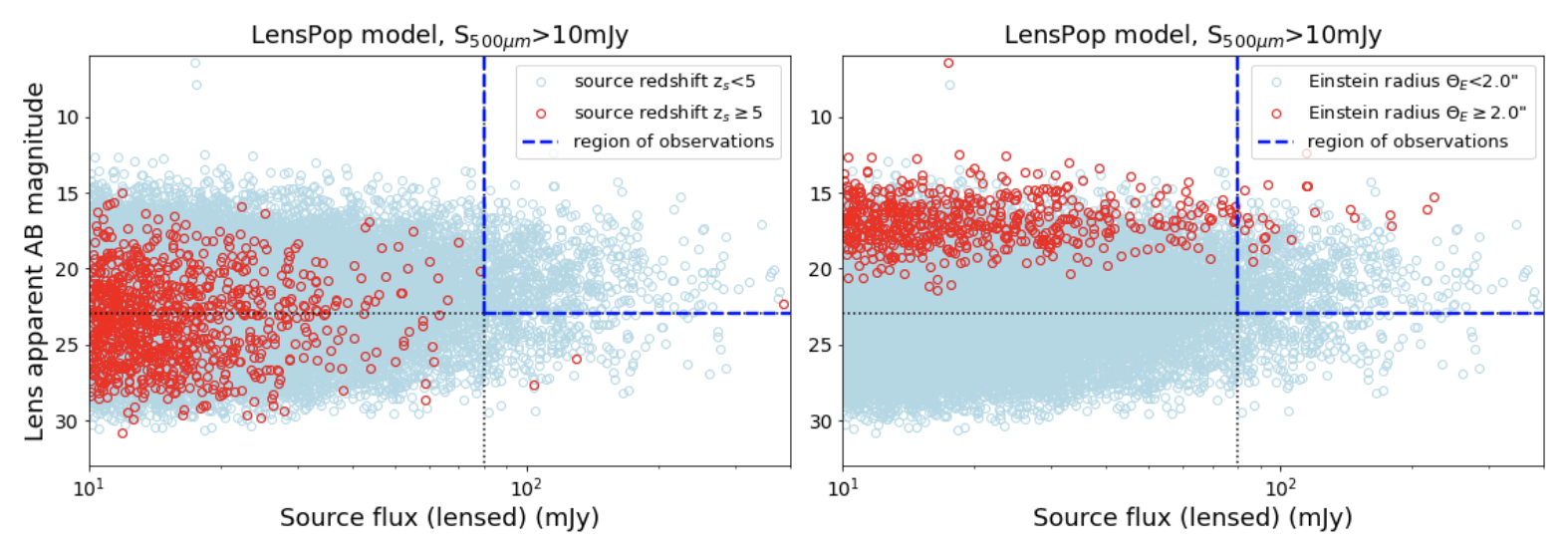}}        
\caption{Observations to date have not covered low-flux sources; according to the model, they have also missed low-flux lenses. The model suggests this means they have also missed (left) high-redshift sources and (right) high Einstein-radii  sources. [Note lens apparent magnitude shown is average of SDSS r, i, and z magnitudes.]}\label{fig:area_zs}
\end{center}
\end{figure*}

Number counts generated by the {\sc LensPop} model show that only 2.6\% of submillimetre lenses have flux $S_{500\upmu\rm m} \geq 80$\,mJy (see Table~\ref{table:counts_by_flux}). Finding lenses at lower flux, however, is much more difficult since the number of bright unlensed sources rapidly overtakes the number of lenses, exceeding them by about three orders of magnitude at $10$\,mJy (see Fig.\,\ref{fig:number_counts2}) where the model predicts that most lenses will be found. The model also predicts that many lower-flux lenses with be at Einstein radii over 2\,arcsec (see Fig.\,\ref{fig:area_zs}, right), making identification with the foreground lens difficult.

Attempts are now being made to define selection criteria to reduce the number of false positives in the search for lenses at lower flux. 

The {\it Herschel} Bright Sources (HerBS) study \citep{Bakx2018} sought to reduce false positives, still using the Negrello criterion at $S_{500\upmu\rm m} \geq80$\,mJy but also setting a minimum to the source redshift ($z_\textrm{S}>$2) to select candidates in the H\nobreakdash-ATLAS fields. The candidates in the Southern Galactic Pole (SGP) field from this study were used in the \cite{Urquhart2022} study discussed above. To see whether this additional selection criterion might bias the results, we have compared the model properties of lenses with $z_\textrm{S}>$2 with lenses at all redshifts, and there does not seem a significant bias in the results. Two of these comparisons (for magnification and lens redshift) are shown in Fig.\,\ref{fig:herbs}.

\begin{figure}
 \begin{center} 
  \resizebox{3.2in}{!}{\includegraphics{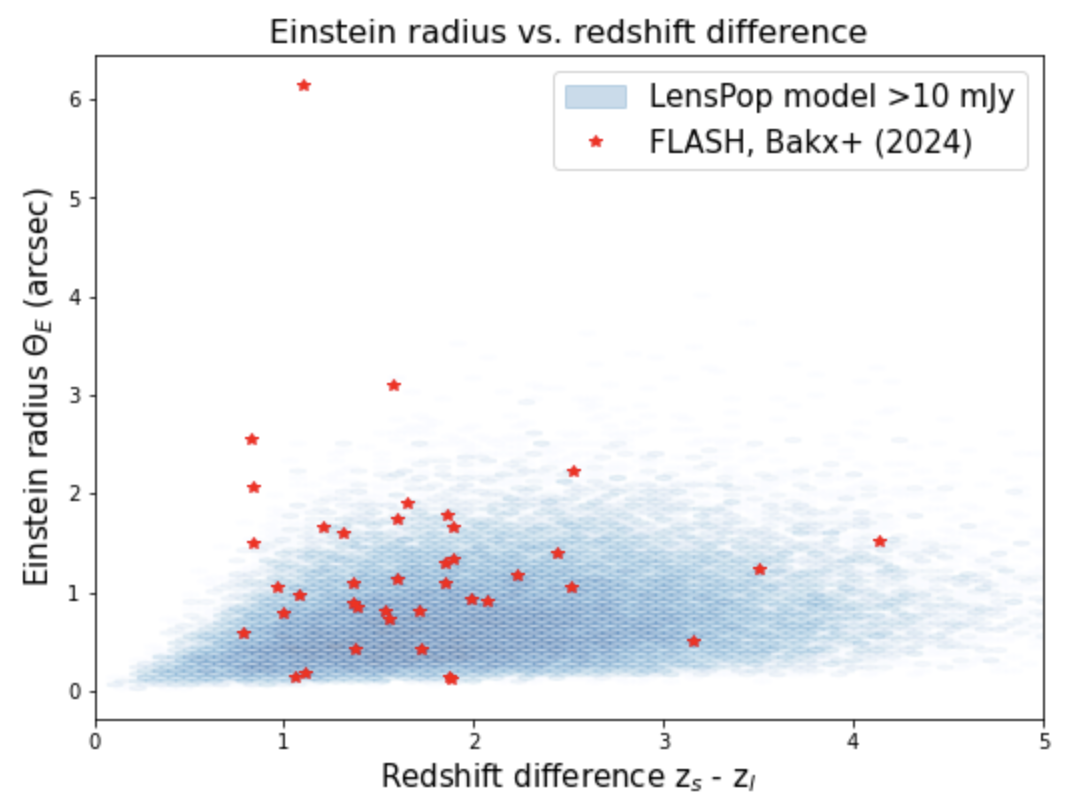}}  
\caption{Comparison between the 40 confirmed lenses in FLASH (\protect\cite{Bakx2024}) and model predictions for $S_{500}>10$\,mJy.}\label{fig:einstein2b}
\end{center}
\end{figure}

A recent study called Faint Lenses from Associated Selection with {\it Herschel} \citep[FLASH;][]{Bakx2024} has identified lenses at lower fluxes ($15$--$85$\,mJy) with ALMA observations, after matching submillimetre sources in the H\nobreakdash-ATLAS equatorial 12h field (area 53.56\,deg$^2$) to nearby sources observed at near-infrared wavelengths in the VISTA Kilo-degree Infrared Galaxy survey \citep[VIKING;][]{Sutherland2012}, which reached a depth of $M_{\ast}>10^8$\,M$_{\odot}$, again with $z_\textrm{S}>2$. The observations with ALMA for 86 sources confirmed that about half (and potentially all) of these candidates are lenses. Our model is consistent their estimate that $\sim$3\,000 further lenses could be discovered by this method in the {\it Herschel} H\nobreakdash-ATLAS catalogues (see Table~\ref{table:counts_by_flux} and Fig.\,\ref{fig:number_counts2}). Matching to the optical SDSS catalogue proved less successful in identifying candidates \citep{Bakx2020}. The number density of confirmed lenses (not candidates) is shown also in Fig.\,\ref{fig:number_counts2}: if the remaining candidates are confirmed as lenses, the density of the FLASH sources would be very close to the model prediction for $S_{500\upmu\rm m}>15$\,mJy. The angular separation of the confirmed lenses is compared to the model predictions of the Einstein radius for lenses $S_{500}>10$\,mJy in Fig.\,\ref{fig:einstein2b}; there are several very high values in the observed data.

Data Release III for H\nobreakdash-ATLAS \citep{Ward2022} also sought to identify near-infrared counterparts in the VIKING survey, in this case for the SGP field, using a matching radius of 15\,arcsec~to H\nobreakdash-ATLAS sources, with the authors estimating a total of 41 lens candidates above 100\,mJy and 5\,923 below 100\,mJy.

A study by \cite{Lammers2022} has used machine learning to identify lens candidates in the {\it Planck} high-redshift catalogue using the flux of all three {\it Herschel}-SPIRE bands, finding new lens candidates embedded in high-density regions and with lower flux.

\section{Discussion}\label{sec:discussion}

\begin{table}
\caption{Potential number of detections of lensed submillimetre sources, and number of sources detectible by NISP but not by VIS (\textquoteleft dropouts'), in the 15\,000\,deg$^2$ of the {\it Euclid} Wide Survey \citep{EuclidWide2022}.} \label{table:euclid_obs}
\begin{center}
\begin{tabular}{|lrrrr|}
\hline
\\

                   & Wavelength       & Limits      & No.  \\\\
\hline

\multicolumn{3}{l|}{\bf Total submm sources:} \\
Whole sky          &  $500\,\upmu$m  & 10\,mJy    &  137\,112      \\    
Euclid Wide Field  &         &                  &  51\,417      \\\\

\multicolumn{3}{l|}{\bf Potential detections by Euclid:}\\
Detections VIS     &  $0.72\,\upmu$m   & 25.0 AB     &   3\,102    \\
Detections NISP (H) &  $1.77\,\upmu$m    & 24.4 AB     &   28\,464 \\\\

\multicolumn{3}{l|}{\bf Dropouts: in NISP, not in VIS:}\\
Redshift $z_\textrm{S}= 1 - 2$     &&&    4\,056  \\   
~~~~~~~~~~~~~~~~~~   $2 - 3$       &&&   16\,677   \\
~~~~~~~~~~~~~~~~~~   $3 - 4$       &&&    4\,438   \\
~~~~~~~~~~~~~~~~~~   $4 - 5$       &&&       186   \\
~~~~~~~~~~~~~~~~~~   $5 - 6$       &&&         5  \\
Total dropouts  &&&   25\,362  \\
\hline
\end{tabular}
\end{center}
\end{table}


%

Our model predicts that the high-flux sources that can be identified by the Negrello criterion account for only about 2.6\% of all submillimetre lenses detected by {\it Herschel} with $500\,\upmu$m fluxes brighter than $10$\,mJy. The most promising avenue to find more lenses in the near term, therefore, would seem to be to look for the lower-flux lenses among the $\sim$million sources in the {\it Herschel} catalogues (see Fig.\,\ref{fig:area_zs}(a) which shows that most high-redshift sources remain to be discovered). The recent FLASH study \citep{Bakx2024} shows a strong promise of identifying many of the lower-flux lenses: their estimate of  $\sim$3\,000 lenses in the {\it Herschel} H\nobreakdash-ATLAS catalogue is consistent with the number predicted by the model. 
 
Comparison with observations to date broadly confirms the accuracy of predictions by the model - if anything, the model under-predicts the number of sources, particularly the highest-magnification sources. One unexpected result from the model was the extent to which very faint lenses generate observable lens systems - lenses with apparent magnitudes at visible wavelengths (the model uses an average of the SDSS r,i,z bands) down to about 28 (see Fig.\,\ref{fig:magnitude2}). The possibility that these ultra-faint lenses might form a sub-population with different properties to the brighter lenses was explored, but there was no indication that these lenses have different properties than the brighter-lens sources in the sample. It does suggest that some apparently unlensed submillimetre sources may be lensed by  faint, undetected lensing galaxies.

\begin{figure*}
 \begin{center} 
 \resizebox{6.5in}{!}{\includegraphics{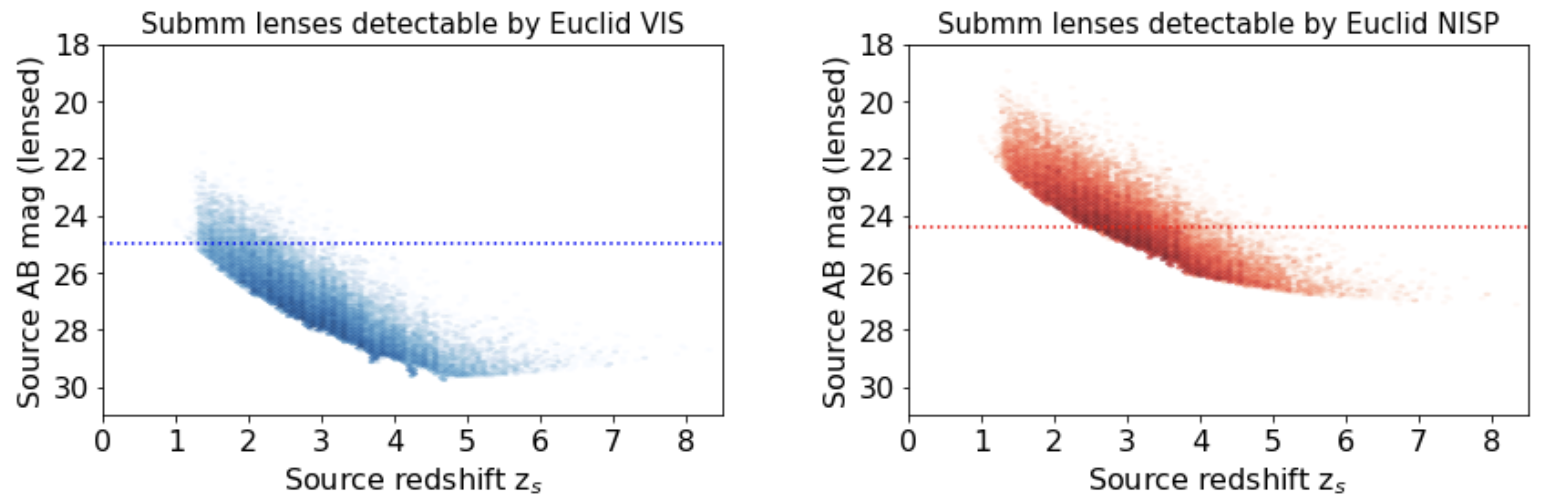}}       
\caption{Potential detections of submillimetre lensed sources by the {\it Euclid} Wide Survey with the VIS and NISP instruments. We have taken waveband centres of $0.72\,\upmu$m for VIS and $1.77\,\upmu$m for NISP (H band). Horizontal lines indicate detection limits for each instrument. The observed flux at $1.77\,\upmu$m is higher than at $0.72\,\mu$m due to the K-correction, assuming the SED of SMMJ2135-0102 (the Cosmic Eyelash) \citep{Swinbank2010} to convert from the $500\,\upmu$m flux.}\label{fig:euclid1}
\end{center}
\end{figure*}

\begin{figure}
 \begin{center} 
 \resizebox{3.3in}{!}{\includegraphics{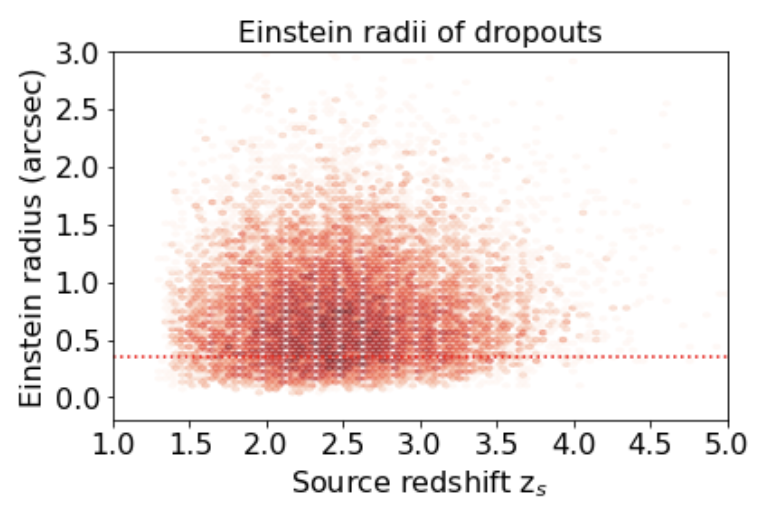}}       
\caption{The model predictions of the Einstein radii of  {\it  Euclid} dropouts - sources detectable by NISP (H filter) but not by VIS (average of SDSS r,i,z filters). The horizontal line shows the limit of the NISP instrument's point-spread-function to separate the source and the lens.}\label{fig:euclid2}
\end{center}
\end{figure}

The Wide Survey \citep{EuclidWide2022} of the {\it Euclid} Space Telescope \citep{EuclidOverview2024} will cover 15\,000\,deg$^2$ and most of the submillimetre lensed sources discovered to date in its field should be detectable (about 1\,360 sources will be detectable for $S_{500\upmu\textrm{m}}>80$\,mJy). The model estimates the VIS instrument \citep{EuclidVIS2024} will be able to detect about 3\,000 of these sources in total, but the NISP instrument \citep{EuclidNISP2024} may detect 28\,000:  the slope of the assumed SED for these sources means a higher flux at NISP wavelengths (see  Table~\ref{table:euclid_obs} and Fig.\,\ref{fig:euclid1}). This results in many sources being detectable in NISP but not in VIS: the model predicts 25\,000 such \textquoteleft dropouts', most of which will be between redshifts 1 to 4 (see Table~\ref{table:euclid_obs}). In many cases, of course, the foreground lens will be detected by VIS, so identifying these dropouts will require a colour or morphological comparison between the sources detected by each instrument. The Einstein radii of these dropouts are shown in Fig.\,\ref{fig:euclid2}. Most of the the foreground lenses are potentially detectable by both instruments. The {\it Euclid} Deep Survey has greater sensitivity, but the much smaller size of its field ($53$\,deg$^2$) will limit its ability to detect more than a handful of lensed submillimetre sources. \cite{Holloway2023} predicted strong lensing frequency in various surveys by using a single empirical catalogue for both source and lens galaxies to identify close galaxy pairs, then applying redshift distributions and similar constraints to those used in {\sc LensPop} to identify which pairs are lens systems. For the {\it Euclid} Wide Area Survey, they predicted 95\,000 lenses detectable by VIS, and 61\,000 by NISP-H (these figures cover all lenses, not just the lensed submillimetre galaxies discussed in this paper).

The ground based surveys by the SPT have proved successful, and they appear to find about one-third the number density of lensed submillimetre galaxies on the sky. The next generation SPT-3G has a sensitivity $\sim5$ times greater than the existing telescope instrumentation, and is planned to survey 1\,500\,deg$^2$. Following that, the CMB-S4 project \citep{Gallardo2022} is planned to be an array of new microwave telescopes at the South Pole and in Chile, and  it is projected to find $\sim10$ times more DSFGs \citep{Reuter2020}. Further candidate lenses may also be identified with the ACT 
\citep{Marsden2014,Gralla2020}.

Ground-based instrumentation mapping speeds in the $450\,\upmu$m atmospheric window are not yet fast enough to duplicate the application of the Negrello criterion to {\it Herschel} surveys, though the SPT and ACT surveys have successfully achieved equivalent selections at longer wavelengths \citep{Reuter2020,Marsden2014}. 
Recent SCUBA-2 surveys found an upturn in number counts of sources $>10$\,mJy at $850\,\upmu$m \citep{Geach2017,Garratt2023}, indicating strongly-lensed galaxies, so a wider-area search could be used to detect lenses. Matching SCUBA-2 data from surveys at $450\,\upmu$m to recent near-infrared data from the {\it James Webb Space Telescope} (JWST) is starting to yield results \citep{Pearson2024}. 

Meanwhile, further developments in millimetre-wave ground-based facilities and detectors offer increasing capabilties for the detection of strongly lensed dusty star-forming galaxies. For example, the TolTEC camera \citep{Bryan2018} has started operations, and uses a new technology to image large areas of the sky at high sensitivity. It is now installed at 
the Large Millimeter Telescope (LMT). There is also the Atacama Large Aperture Submillimeter Telescope \citep[AtLAST;][]{Klaassen2020}, a 50-metre class single dish observatory, which is currently under design and planned to start observing in the mid-2030s. 

NASA has selected the PRIMA far-infrared mission as one of two missions for further study in its Probe-class program, one of which is planned for launch in the 2032. In the L2 orbit, this should provide synergies with the AtLAST and upgraded ALMA telescopes, the far-infrared region being essential to fully explore questions such as the connections between the growth of galaxies and black holes, and the evolution of metallicity. Greater depth and sensitivity of  observations, particularly of bright lensed submillimetre sources, will enable the study of the ISM properties of the sources, and high-redshift lenses will enable the investigation of dark matter halo substructure, to provide an important test of current cosmological assumptions.
Understanding the wide variation in number density being found in different fields could also contribute to the justification for a wide-field survey to study a larger area of the sky in such a mission.

\section*{Data availability}

All the catalogue data used and referenced in this paper are publicly available in the papers cited. The adapted {\sc LensPop} code presented in this paper is available at \url{https://github.com/chrissedgwick/LensPop\_submm}~.


\section*{Acknowledgments}

Thanks to Tom Collett for making the {\sc LensPop} model available and for helpful comments. SS thanks STFC for support under grants ST/P000584/1 and ST/S006087/1. We also thank the anonymous referee for helpful comments.


\bibliography{references}
\bibliographystyle{mnras}

\bsp
\label{lastpage}

\end{document}